\documentclass{IEEEcsmag}

\usepackage{upmath}
\usepackage{soul}
\usepackage[most]{tcolorbox}
\usepackage{tikz}

\usepackage{multirow}
\usepackage{subcaption}
\usepackage{graphicx}

\usepackage{multirow}

\usepackage{lineno}

\usepackage{xargs}                      
\usepackage{enumerate}

\usepackage[colorinlistoftodos,prependcaption, textsize=tiny,textwidth=1.8cm]{todonotes}
\newcommandx{\improvement}[2][1=]{\todo[linecolor=blue,backgroundcolor=blue!25,bordercolor=blue,#1]{#2}}

\usepackage[colorlinks,urlcolor=blue,linkcolor=blue,citecolor=blue]{hyperref}

\usepackage{lscape}
\usepackage{soul}

\usepackage{adjustbox}

\usepackage{pgfplots}

\usepackage{pgfplotstable}

\jvol{XX}
\jnum{XX}
\paper{8}
\jmonth{January}
\jname{IEEE Security \& Privacy}
\pubyear{2023}

\begin{document}



\title{Supporting AI/ML Security Workers through an Adversarial Techniques, Tools, and Common Knowledge (AI/ML ATT\&CK) Framework}

\author{Mohamad Fazelnia}
\affil{Global Cybersecurity Institute (GCI), Rochester Institute of Technology (RIT), USA}

\author{Ahmet Okutan}
\affil{Global Cybersecurity Institute (GCI), Rochester Institute of Technology (RIT), USA}

\author{Mehdi Mirakhorli}
\affil{Global Cybersecurity Institute (GCI), Rochester Institute of Technology (RIT), USA}


\begin{abstract}
This paper focuses on supporting AI/ML Security Workers\textemdash~ professionals involved in the development and deployment of secure AI-enabled software systems. It presents AI/ML Adversarial Techniques, Tools, and Common Knowledge (AI/ML ATT\&CK) framework to enable AI/ML Security Workers intuitively to explore offensive and defensive tactics. 

\end{abstract}

\maketitle
\chapterinitial{Introduction} In recent years, there has been a significant increase in the use of artificial intelligence and machine learning (AI/ML) in a variety of application domains, ranging from autonomous driving cars to medical diagnosis tools. Such increased usage of AI/ML techniques in real-world applications attracted adversaries and cyberattackers to exploit the weaknesses in these techniques to achieve malicious goals.  AI/ML capability builders need to employ appropriate mitigation techniques during the development and maintenance stages to minimize the attack surface of the AI-software products~\cite{tabassi2019taxonomy}. This is a challenging task considering the variety of AI/ML techniques, their unique vulnerabilities, and the emergence of new attack vectors. 

While the shortage of skilled information security workers continues to grow, this problem is exacerbated when engineering AI-Enabled software systems — for the simple reason that it requires security workers who are AI/ML subject matter experts, and demand for such expertise continues to exceed the supply. This paper aims to provide a framework that can be used by AI engineers to transform them into \textit{AI/ML Security Workers}.

To develop  AI/ML-enabled systems resilient against cyberattacks, AI engineers must be aware of the existing attacks and their distinctive characteristics, such as the underlying techniques, evolving scenarios, and goals. For classical software systems, there are extensive resources such as MITRE's CWE, CAPEC and ATTCK frameworks that enumerate software weaknesses and common attacks~\cite{attck, CWE}, however these frameworks do not cover the security issues of AI/ML systems. Having similar resources for AI engineers would enable them to become AI/ML Security Workers by guiding them in investigating potential attack scenarios against their AI products and choosing appropriate defensive measures. While existing resources such as NIST Taxonomy and Terminology of Adversarial Machine Learning~\cite{tabassi2019taxonomy} and other similar work~\cite{10.1145/2046684.2046692, 9099439} provide a common language to discuss various types of attacks using an upper taxonomy, they often lack details and concrete attack/defense techniques necessary to make actionable decisions for a given AI system~\cite{10.1145/3442167.3442177}. 
To fill this gap we need to support the training of the next generation of AI/ML Security Workers, and ease their day-to-day tasks. To this end, we argue that it is necessary to have an \textit{intuitive framework} that guides AI/ML Security Workers, particularly the novices to ensure that cybersecurity thinking is explicitly integrated in their daily AI engineering and decision-making activities.

This paper, therefore, discusses \textbf{AI/ML Adversarial Techniques, Tools, and Common Knowledge (AI/ML ATT\&CK framework)}, an intuitive and comprehensive framework based on a systematic literature review (SLR) that characterizes the \textit{offensive} and \textit{defensive} techniques, tactics, and tools related to the security of AI/ML-enabled systems. AI engineers can employ the AI/ML ATT\&CK framework during software design and implementation as an easy-to-use decision support system to identify the weaknesses associated with the AI models, explore viable attacks, and consider appropriate mitigation techniques and security measures. Our contributions are four-fold:
\begin{itemize}
\item We provide a definition for \textit{AI/ML Security Workers}, describing their role as AI cybersecurity professionals and advocate individuals in such role on taking both \textit{offensive} and \textit{defensive} measures. 
\item We provide an extensive knowledge base of 102 \textit{attacks}, 65 \textit{mitigation techniques}, and 105 \textit{tools} related to the security of AI/ML-enabled software systems. This catalog is created based on an SLR to categorize and characterize currently available attack scenarios, mitigation techniques, and tools discussed in 860 papers.
\item We define an AI/ML ATT\&CK framework that aims to support both \textit{offensive} and \textit{defensive} AI/ML Security Workers. The \textit{Offensive View} is organized as a decision tree, which enumerates 26  \textit{attack scenarios}, and is mapped to 102 concrete attack techniques and existing tools capable of implementing the attacks. The \textit{Defensive View} enumerations 25 \textit{defense scenarios} and is mapped to 65 mitigation techniques.


\item An online interactive version of the framework is provided to support AI/ML Security Workers, especially less-experienced ones. A limited scope user study has demonstrated promising results on the effectiveness of such framework.
\end{itemize}

All research materials, the generated \textbf{AI/ML ATT\&CK} framework, developed  attack mitigation, datasets, and investigated tools are shared with the community at \href{http://www.aimlattack.com/}{\textcolor{blue}{www.aimlattack.com}}.


\begin{table*}[!ht]
\caption{Review Protocol}
\label{tab:Protocol}
\begin{tabular}{ |p{2.3cm}|p{12cm}| } 
\hline
\textbf{Research Questions} & (1) What are the AI/ML-specific weaknesses and vulnerabilities and how do attackers use them to exploit the system and execute organized attacks?

(2) What are the potential mitigation techniques that can be leveraged to prevent and mitigate the vulnerabilities and improve the system’s robustness?  \\ \hline
\textbf{Dates} & 2000-2021  \\ \hline
\textbf{Databases} & IEEE Xplore Digital Library, ACM Digital Library, \\ \hline
 
\textbf{Search Criteria} & English, Search in title, abstract and keywords.  \\ \hline
\textbf{Search Keywords} & \{(\textit{Artificial Intelligence} \textbf{OR} \textit{Machine Learning} \textbf{OR} \textit{Deep Learning} \textbf{OR} \textit{Neural Networks}) \textbf{AND} (\textit{threat} \textbf{OR} \textit{mitigation} \textbf{OR} \textit{adversary})\}.  \\ \hline
\textbf{Inclusion/Exclusion Criteria} & Inclusion: Full Paper, Focus on Attacks and Weaknesses on AI/ML Models, Focus on Improving the Robustness Against Attacks; Exclusion: Not Written in English; Reports, Abstract, Ideas, Summaries and Discussions; Duplicated Studies. \\ \hline
 
\textbf{Number of Papers} & Initial results \textit{20321}; After First Review Step: \textit{3329}; Fully synthesized papers: \textbf{\textit{860} papers}. \\
\hline
\end{tabular}
\vspace{-10pt}
\end{table*}

\section{AI/ML Security Workers}
\label{sec:Workers}
Who are AI/ML Security Workers? We define them as cybersecurity professionals~\cite{Julie,8805749,8804457} who contribute to the development of secure AI-enabled software systems through monitoring the training and deployment of AI systems, as well as formulating new adversarial attack scenarios, measuring their severity, and devising novel defense strategies for the robustness and integrity of AI-enabled systems in different application domains. AI/ML security workers are responsible for implementing, validating, justifying, and advocating the required defenses and mitigation strategies. While this role is not yet defined in standardized Work Roles such as NIST's Workforce Framework for Cybersecurity (NICE)~\cite{NICE}, recent job posts demonstrate a growth in the demand for \textit{AI security engineers}, \textit{Adversarial AI} experts, and many other roles with similar responsibilities.
According to a recent publication~\cite{bughin2018skill}, just about every industry needs workers with AI skills as they focus on technologies to give computers the capability to think, learn, and adapt. 
To utilize AI at its optimal potential, AI engineers need to have programming skills and knowledge in statistical learning, data modeling, data evaluation, software engineering and system design, distributed computing, and in general, the ability for conceptual thinking to understand how a product is used and how it can be used more effectively. 
We argue that AI/ML Security Workers, in addition to the above skills, need to be equipped with \textit{offensive} and \textit{defensive} thinking so they can identify potential \textit{attack scenarios}, \textit{abuse cases}, as well as \textit{mitigation techniques} as they engineer AI Software. 
AI/ML Security Workers will deal with complex tasks that require creativity, critical thinking, complex information processing and decision-making. It has been demonstrated that such tasks need higher cognitive skills~\cite{bughin2018skill}. Unlike software security workers who have access to many resource such as OWASP Top 10~\footnote{OWASP Top 10: https://owasp.org/Top10/}, Security Checklists, Secure Coding Practices, IEEE Top 10 Design Flaws~\footnote{IEEE Top 10 Software Security Design Flaws: https://cybersecurity.ieee.org/blog/2015/11/13/avoiding-the-top-10-security-flaws/} and more, AI/ML Security Workers currently lack comprehensive resources to support them in decision-making. In particular, there are no checklists or other frameworks that enumerate AI/ML weaknesses and attack scenarios that can support both novices and experts in systematically identifying known flaws in AI Software.
Despite the significance of security in AI software, there is a lack of academic research, especially regarding any holistic guidelines to support AI/ML Security Workers.
Our work aims to leverage SLR with the objective of integrating and synthesizing existing knowledge to provide new insights for AI/ML Security Workers. 
This work aims to characterize AI/ML cybersecurity \textit{techniques}, \textit{tactics}, and \textit{tools} in a way that can help AI/ML Security Workers better understand the attack surface of the AI-powered systems, reason about the AI-specific vulnerabilities in the system, and employ the best defense techniques to mitigate the vulnerabilities. 
In particular, we aim to support two groups of AI/ML Security workers who investigate the following questions in their day-to-day activities.

\textbf{Offensive AI/ML Security Workers:} What are the AI/ML-specific weaknesses and vulnerabilities, and how do attackers use these to exploit the system and execute organized attacks?

\textbf{Defensive AI/ML Security Workers:} What are the potential mitigation techniques that can be leveraged to prevent and mitigate the vulnerabilities and improve the system's robustness?

\section{Methodology} \label{sec:methodology}

To create such a comprehensive AI/ML ATT\&CK framework to serve as a guideline, we performed an SLR following the guidelines provided by Kitchenman \cite{kitchenham2004procedures}.

\subsection{Review Protocol}
\label{sec:2}

Table~\ref{tab:Protocol} summarizes our SLR protocol. We performed the SLR in two of the most popular computer science archives; IEEE Xplore and ACM Digital Library.

First, to find the appropriate search keywords, we collected an initial set of relevant works by manual exploration and then searched for the common and relevant words in their titles and abstracts. Then, as we found new relevant papers, we added them to the collection and refined the set of keywords. 
Based on this procedure, we formed the following search query: \{\textit{(Artificial Intelligence} OR \textit{Machine Learning} OR \textit{Deep Learning} OR \textit{Neural Networks}) AND (\textit{threat} OR \textit{mitigation} OR \textit{adversary})\}. We also used other queries that relied on the variations of the terms. The variations consist of the words with and without the possible prefix and suffix for each keyword (e.g., adverse, adversarial, adversary, etc.) This review was conducted on peer-reviewed papers published between the years 2000 and 2021. 

After extracting the papers from IEEE Xplore and ACM digital libraries, we documented the results to prepare them for the review process. From the initial search, we extracted \textit{20321} papers. Then, we checked the paper's titles and filtered out the papers based on the inclusion and exclusion criteria represented in Table~\ref{tab:Protocol}. After the first filtration process, we gathered \textit{3329} papers. Then, we applied the inclusion and exclusion criteria on the abstracts of the remaining papers. At the end of this step, we gathered \textit{860 papers}. Finally, we went through the papers and started the detailed literature review. 

After reviewing the first 10\% randomly selected papers, the initial skeleton of the AI/ML ATT\&CK framework was formulated, consisting of the framework structure and the proposed attributes to model the cyber adversaries and security techniques~\cite{workshop-paper}, as well as the related tools and toolchains. Along with reviewing the papers and adding new techniques to the framework, we updated the framework skeleton, until all the papers were reviewed.
Moreover, during reviewing each paper, we performed a \textit{snowballing} process to review other related works cited in the paper. 
As a result, we present a unifying taxonomy of AI/ML-specific cyberattacks and cyberdefenses that provides the information that can help AI/ML capability builders to deploy appropriate cybersecurity tactics during system design to deliver a reliable and secure AI-based system.

\section{AI/ML Adversarial Techniques, Tools \&  Common Knowledge (AI/ML ATT\&CK) Framework}
\label{sec:Framework}

Based on the conducted literature review and analyzing the papers described in the previous section, we developed our AI/ML ATT\&CK Framework. 
We organized the framework around three main components: \textit{Attack}, \textit{Mitigation}, \textit{Tool}, and two views: \textit{Offensive} and \textit{Defensive}.
The three components of our framework and their intended usage by AI/ML security workers is shown in Figure~\ref{fig:compts}. The \textit{Offensive} and \textit{Defensive} views provides the AI/ML Security Workers with scenario-based decision trees to identify appropriate techniques and tools for their offensive/defensive tasks. 

The three main components of the framework are explained as follows:

\begin{figure}[!tp]
\vspace{-5pt}
    \centering
    \includegraphics[width=0.47\textwidth]{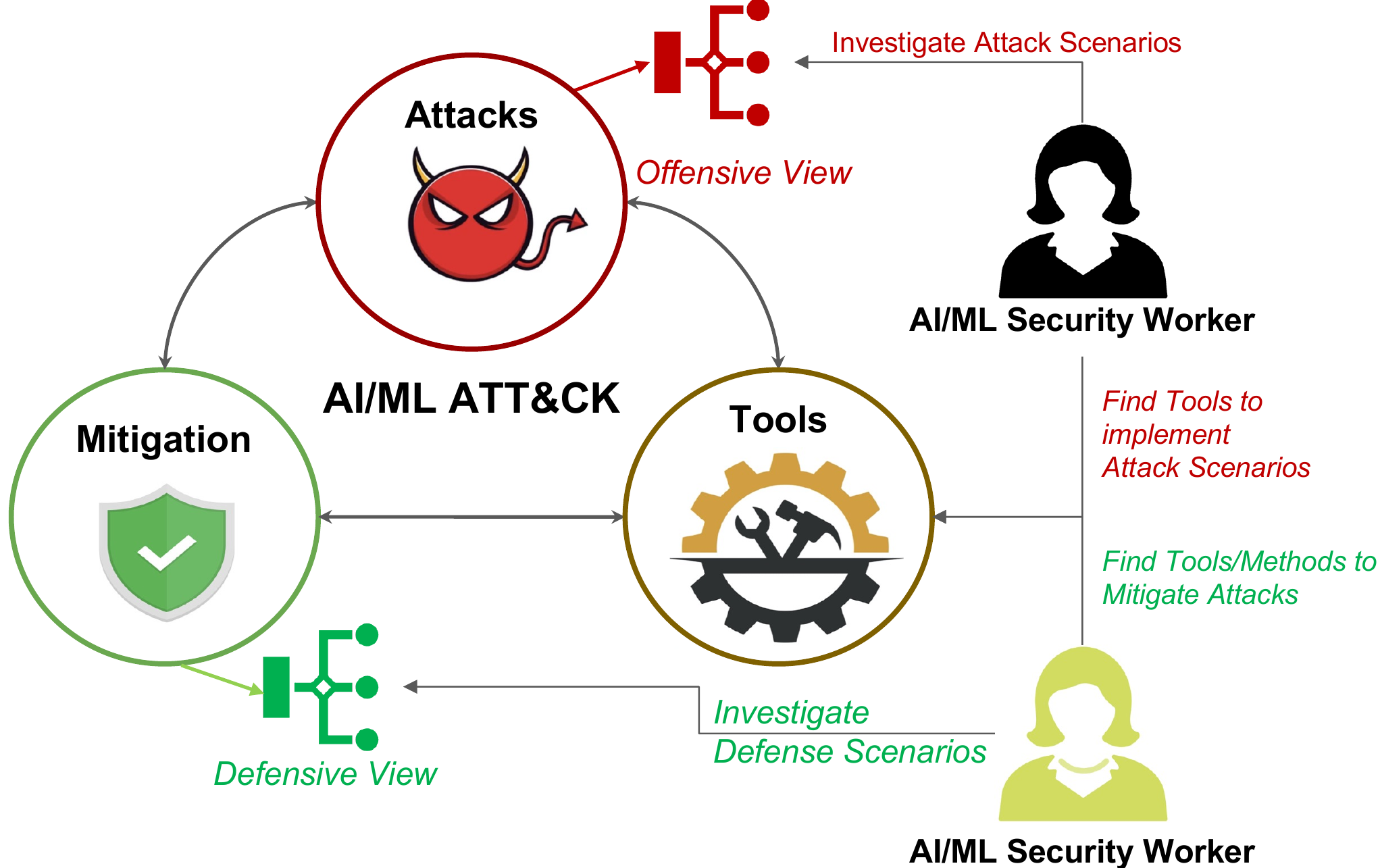}
    \caption{The framework consists of three main components, \textit{Attacks}, \textit{Mitigations}, and \textit{Tools}. The framework connects and unites these three components to help AI/ML Security Workers identify related cybersecurity techniques. The \textit{Offensive} and \textit{Defensive} scenarios help Security Workers identify and select appropriate cybersecurity techniques.}
\label{fig:compts}
\end{figure}

\begin{enumerate}
    \item  \textit{Attack:} Represents a comprehensive analysis of the offensive techniques against AI/ML-enabled systems, as well as the adversary's goal, assumptions, and capabilities. 
    \item \textit{Mitigation:} Provides a detailed analysis of the defensive techniques, their effects, approach, and applicability of offensive techniques. 
    \item \textit{Tool:} Illustrates tools and toolchains capable of providing the offensive and defensive techniques for AI/ML systems.
\end{enumerate}

Next section explores each component and describes the features in detail.
\begin{figure}[!htbp]
\vspace{-5pt}
    \centering
  \includegraphics[width= 0.51\textwidth]{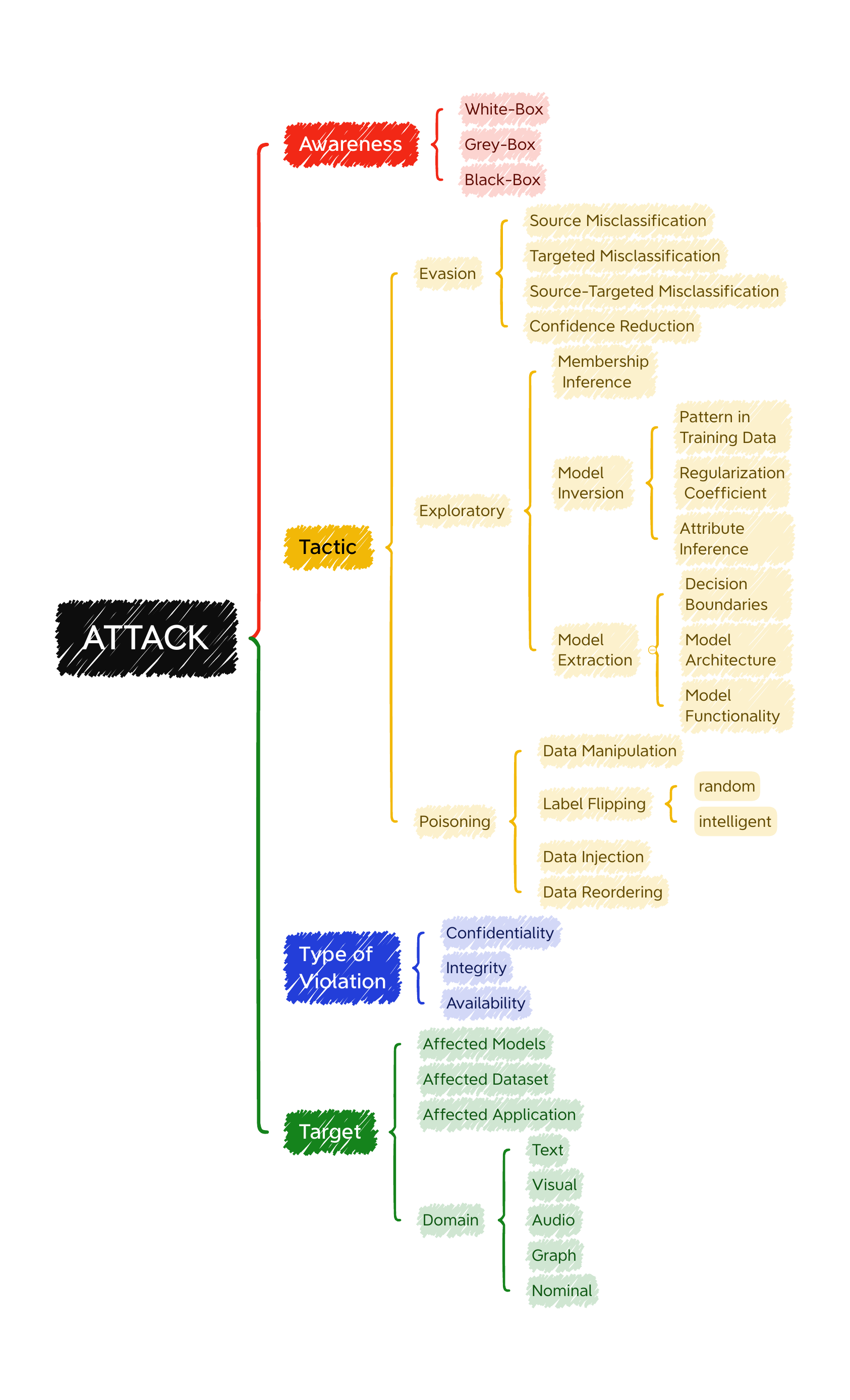}
\vspace{-20pt}
\caption{Characterization of AI/ML-specific offensive techniques. Each leaf node of this tree, in the online framework, expands to concrete offensive techniques for further investigation.}
\label{fig:attack}       
\vspace{-10pt}
\end{figure}
Current version of AI/ML ATT\&CK framework contains more than 100 attack techniques, 65 mitigation techniques, and 105 tools related to the security of AI/ML techniques. Each entry in this framework provides a descriptive analysis of the technique as well as the corresponding offensive or defensive techniques. 
Moreover, the online framework provides information about the dataset used in each technique and its characteristics.

\subsection{AI/ML ATT\&CK Components}
\label{components}
This section provides a description of components, their features, and relationships.
\subsubsection{Attacks}
\label{sec:8} represents a detailed analysis of the offensive techniques against AI/ML-powered systems, as shown in Figure~\ref{fig:attack}. Each part of this component is described as follows:

\textbf{Awareness:}
Some cyberattacks require information such as training data, models, and architecture, while other attacks can be executed without such knowledge. This feature represents the level of information that is accessible to the adversary as follows:

\begin{enumerate}
    \item \textit{White-Box:} The attacker has full knowledge of the target model and the training data.
    \item \textit{Grey-Box:} The attacker has partial knowledge about the model and the training data.
    \item \textit{Black-Box:} The adversary has no knowledge about the model or the data.
\end{enumerate}

\textbf{Tactic:} This component represents adversary's approach, which is categorized as follows:

\begin{enumerate}
    \item \textit{Poisoning:} The adversary tries to undermine the learning process during model training to manipulate the system's behavior during inference time. In this attack, the attacker poisons the dataset through the following approaches: 
    \begin{itemize}
        \item \textit{Data Manipulation:} The attacker manipulates the existing training samples that results in wrong output during inference time.
        \item \textit{Label Flipping:} By adjusting the label of the training samples, the attacker aims to interrupt the learning process in two ways of ${1)}$ \textit{random} and ${2)}$ \textit{intelligent}.
        \item \textit{Data Injection:} The attacker injects malicious samples in to the training dataset that corrupts the learning process.
        \item \textit{Data Reordering:} The attacker changes the order of the data feeding in training to corrupt the learning process.
    \end{itemize}
    \item \textit{Evasion:} The attacker aims to fool the ML system to misbehave by modifying the legitimate input to result in wrong predictions during inference time. Depending on the input's original label and the target class, this attack is categorized as follows~\cite{7467366}:
    \begin{itemize}
        \item \textit{Confidence Reduction:} The attacker tries to reduced the confidence scores of the predicted classes.
        \item \textit{Source Misclassification:} The adversary tries to change the prediction to any label other than the correct one. 
        \item \textit{Targeted Misclassification:} The attacker aims to force the model to misclassify any given input to the specified class.
        \item \textit{Source-Targeted Misclassification:} The attacker forces the model to misclassify a certain set of inputs to a specific class.
    \end{itemize}
    \item \textit{Exploratory}: The adversary aims to explore the system's private information, such as training data, learning algorithm, and the decision boundaries. There are three types of attacks based on the stolen information:
    \begin{itemize}
        \item \textit{Membership Inference:} The adversary aims to determine whether a specific sample has contributed to the training process.
        \item \textit{Model Inversion:} The adversary aims to reconstruct the training data and the samples representing each class of the training dataset.
        \item \textit{Model Extraction:} The adversary aims to extract the model, the decision boundaries, or build a functionally similar model to the target model.
    \end{itemize}
\end{enumerate}

Each of the above tactics is further classified into more granular classes, as shown in Figure~\ref{fig:attack}, and is accessible to AI/ML Security Workers for deeper investigations available at 
\href{http://www.aimlattack.com/}{\textcolor{blue}{online AI/ML ATT\&CK framework}}.

\textbf{Type of Violation:}
This feature represents the type of violation caused by the adversary, which can be one or combination of the following groups: 
\begin{enumerate}
    \item \textit{Confidentiality:} The adversary aims to extract system's private information.
    \item \textit{Integrity:} The adversary forces the system to generate inaccurate results by processing malicious samples.
    \item \textit{Availability:} The adversary aims to corrupt the normal behavior of the system for legitimate inputs.
\end{enumerate}

\textbf{Target:} Represents the entities that are vulnerable to the attack as follows:

\begin{enumerate}
    \item \textit{Affected Models:} Represents the ML and Deep Learning (DL) models that are vulnerable to the adversary.
    \item \textit{Affected Applications}: Illustrates the real world applications, services, software and tools that can be affected by the corresponding adversary.
    \item \textit{Affected Dataset}: Represents the dataset on which the attack is carried out.
    \item \textit{Domain}: Represents the type of the data on which the attack is carried out, which can be in the form of \it{text, visual, audio, graph, or nominal.}
\end{enumerate}

\subsubsection{Mitigation}
\label{sec:9} represents a detailed analysis of the AI/ML-specific defensive techniques, as shown in Figure~\ref{fig:2}. Each component of this framework is described as follows:

\begin{figure}
  \includegraphics[width=0.52\textwidth]{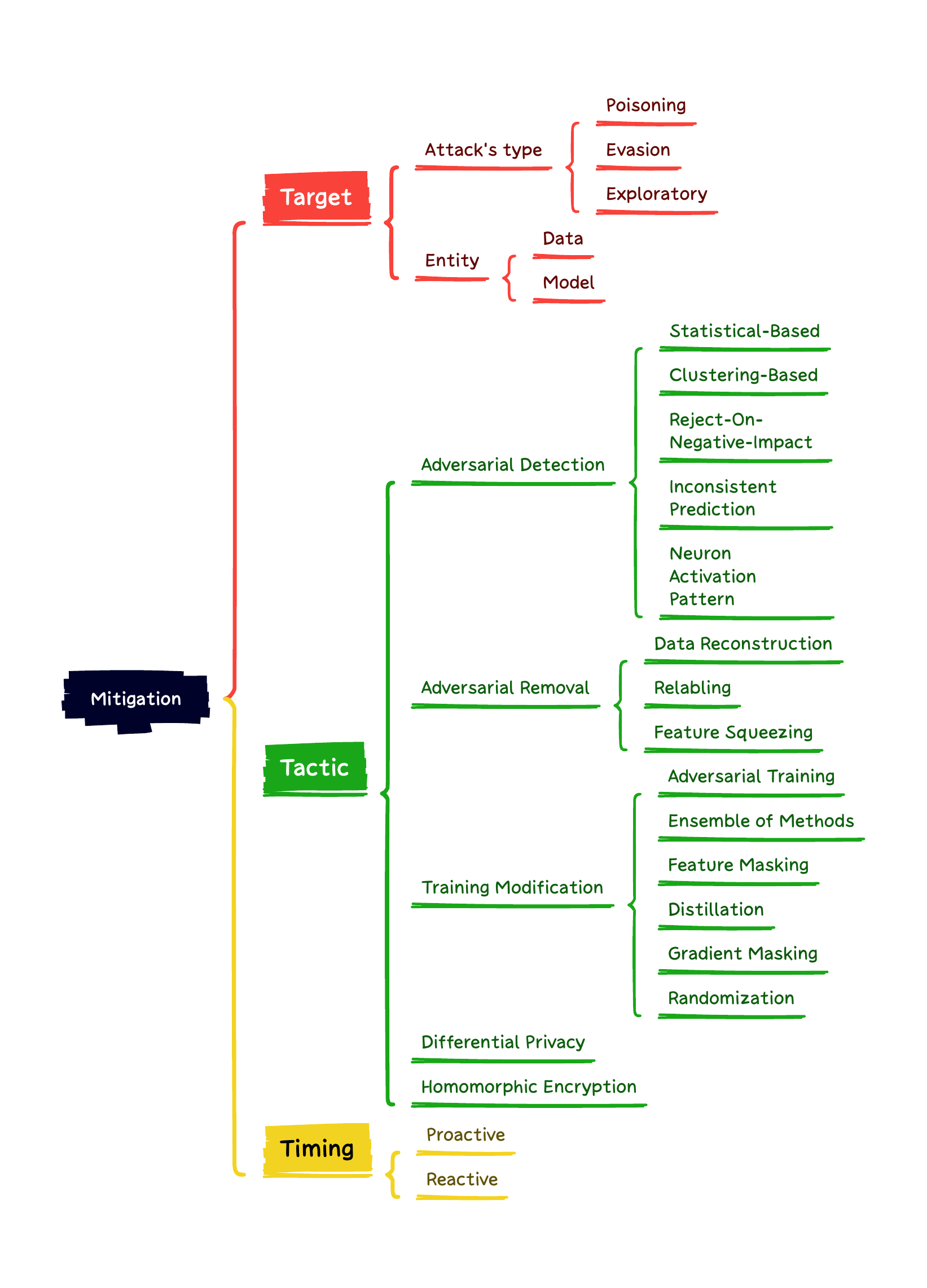}
\vspace{-20pt}
\caption{Categorization of the Mitigation techniques. Each leaf node of this tree on the website expands to the concrete techniques identified in the literature. It can be used to explore defensive techniques and understand their characteristics.}

\label{fig:2}       
\vspace{-10pt}
\end{figure}

\textbf{Target:} Represents the corresponding attack and the affected \textit{data} and \textit{model} entities. 

\textbf{Tactic:} This component classifies the mitigation technique based on the approach as follows: 
\begin{enumerate}
    \item \textit{Adversarial Detection:} Detects malicious samples to prevent evading into the system.
    \item \textit{Adversarial Removal:} Removes the perturbations from the data and generates clean data.
    \item \textit{Training Modification:} Modifies the training process to make the model more robust against attacks.
    \item \textit{Differential Privacy:} Protects the privacy of the system by sharing patterns of the training dataset while withholding information about the individuals in the training dataset.
    \item \textit{Homomorphic Encryption:} Provides an encrypted form of the data for training the model.
\end{enumerate}

Each of the above tactics is further categorized into more granular classes, as shown in Figure~\ref{fig:2}, and is accessible to AI/ML security Workers for deeper investigations available at \href{http://www.aimlattack.com/}{\textcolor{blue}{online AI/ML ATT\&CK framework}}.

\textbf{Timing:} There are two stages when the mitigation techniques can be deployed:
\begin{enumerate}
    \item \textit{Proactive:} The mitigation technique should be deployed before or during the model training.
    \item \textit{Reactive:} The mitigation technique should be deployed after model training.
\end{enumerate}

\subsubsection{Tool}
\label{sec:10}
characterizes the tools and toolchains capable of deploying offensive and defensive techniques related to the security of AI/ML systems as shown in Figure~\ref{fig:3}. Each part of this component is explained as follows:
\begin{figure}

  \includegraphics[width=0.51\textwidth]{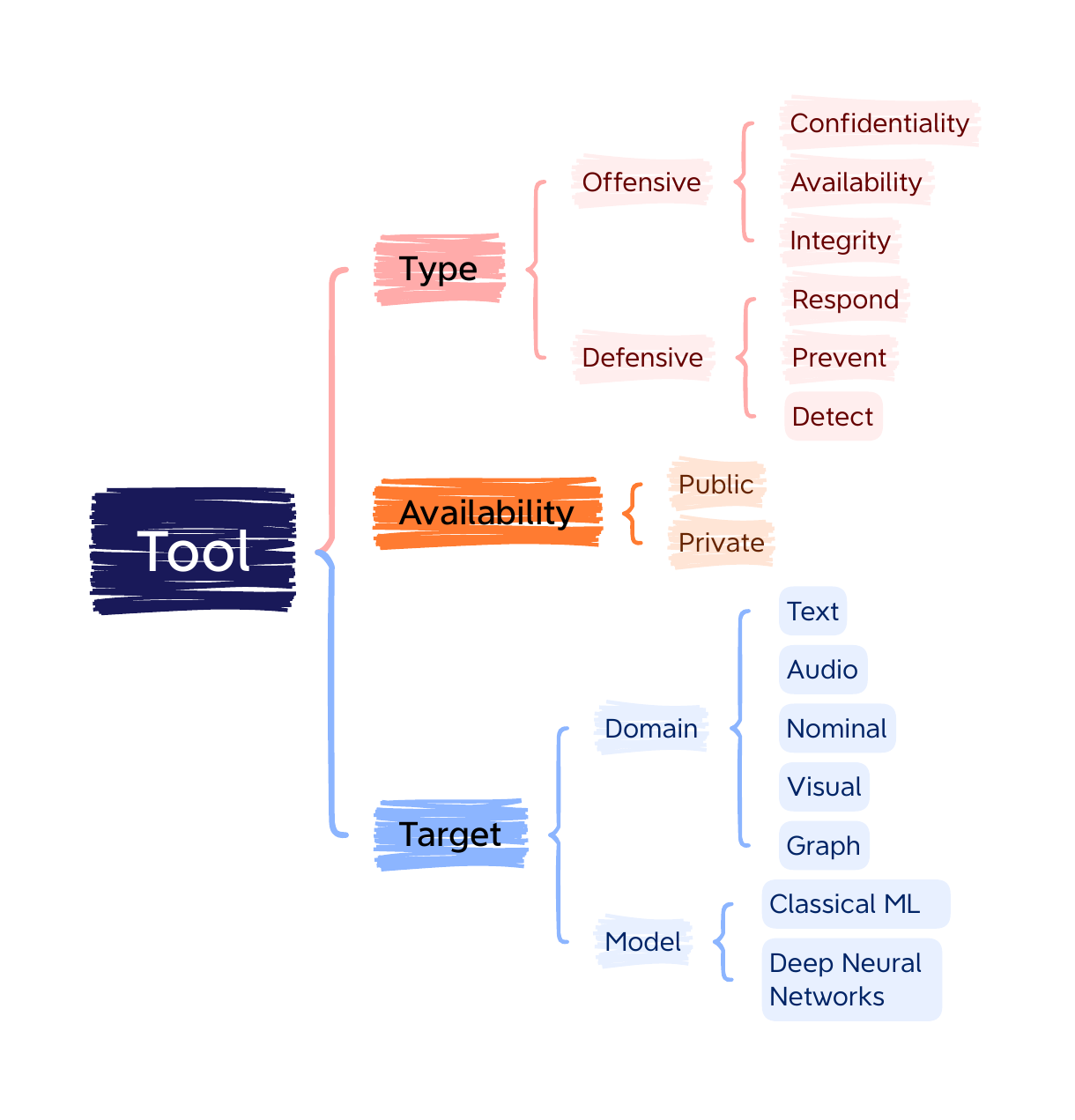}
  \vspace{-20pt}
\caption{Tool component. It characterizes tools and toolchains capable of implementing the offensive and defensive techniques. On the website, each leaf node of this tree expands to show the concrete tools and its code repository as identified in the literature.}
\label{fig:3}       
\vspace{-5pt}
\end{figure}

\textbf{Type:} Represents whether the tool is offensive or defensive. 

\textbf{Availability:} Illustrates whether the tool is publicly available or private.

\textbf{Target:} Explains the data and model entities that are required to deploy the tool.

\begin{figure*}[!ht]
\vspace{-5pt}
\hspace{-10pt}
    \includegraphics[width=1.04\textwidth]{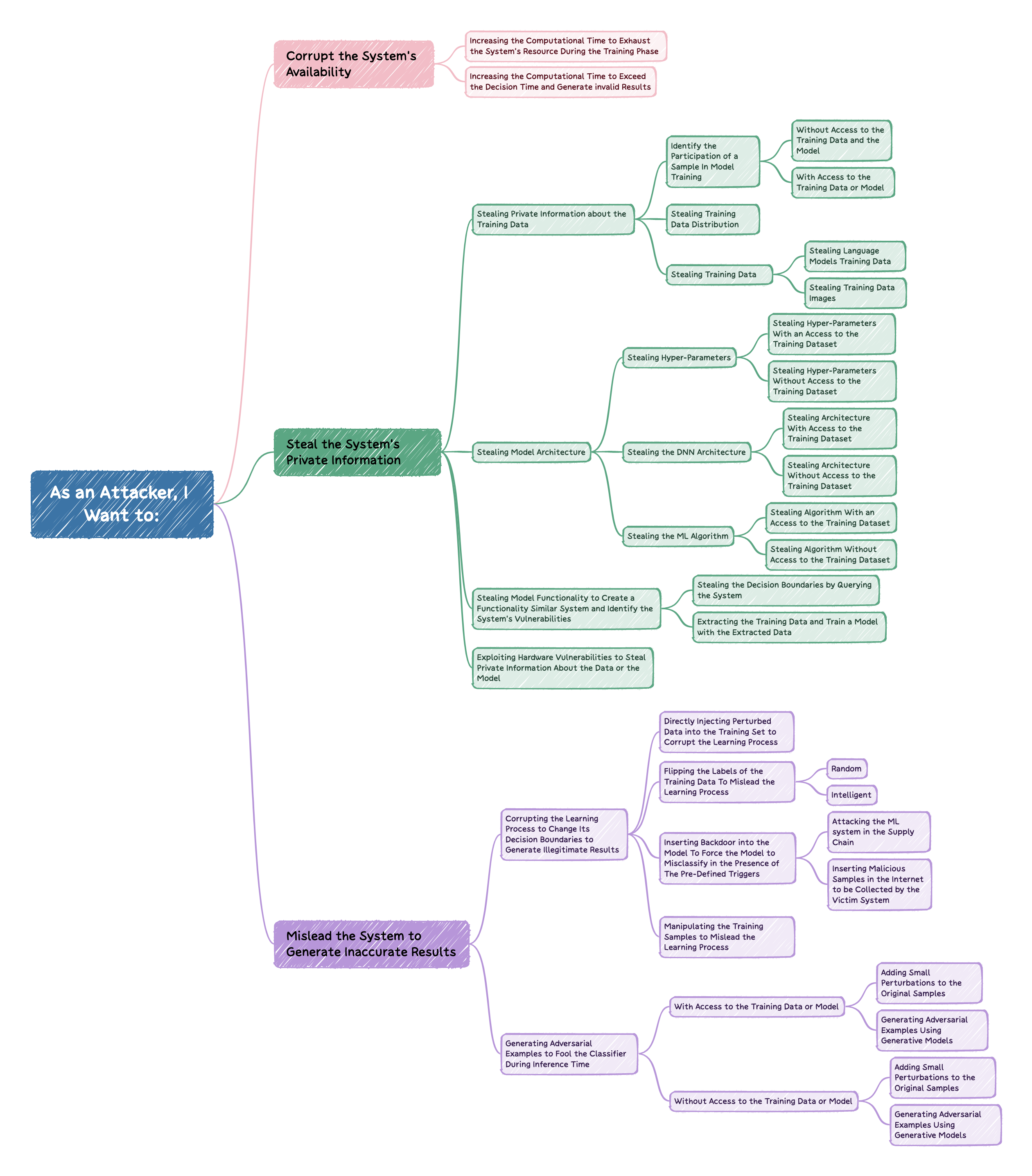}
        \vspace{-20pt}
    \caption{\textbf{Offensive View of AI/ML ATT\&CK Framework.} It provides an overview of the attack scenarios from the attacker's perspective. Offensive AI/ML Security Workers follow the provided step-by-step guidance to identify the related attacks and the tools. In the website, the leaf nodes of this view are connected to the concrete attack techniques for further investigation.}
    \vspace{-10pt}
    \label{fig:offensive}
\end{figure*}

The AI/ML ATT\&CK provides features to characterize the offensive and defensive methodologies related to the security of AI/ML techniques. The framework enables AI/ML Security Workers to understand the attack surface of different AI/ML techniques. Moreover, it represents appropriate cyberdefense techniques to build robust systems that withstand various threats. To our knowledge, this is the first study that thoroughly characterizes tools and toolchains related to AI/ML security and can practically guide the AI/ML Security Workers to choose the best tools capable of implementing offensive and defensive tactics.

\section{Supporting AI/ML Security Workers}
\label{sec:Scnearios}
Simply providing access to a large number of AI/ML robustness techniques cannot fully support the users in choosing the best techniques to deploy. They need easy-to-use guidance to explore various techniques, identify the best approach, and reason about their choices. 
To this end, two additional views, \textit{Offensive View} and \textit{Defensive View,} have been compiled in the AI/ML ATT\&CK framework.
These views provide decision trees representing Offensive and Defensive Scenarios that can guide the AI/ML Security Workers to investigate the system's robustness from the attacker and defender perspectives.

\begin{figure*}
  \vspace{-10pt}
    \includegraphics[width=0.94\textwidth]{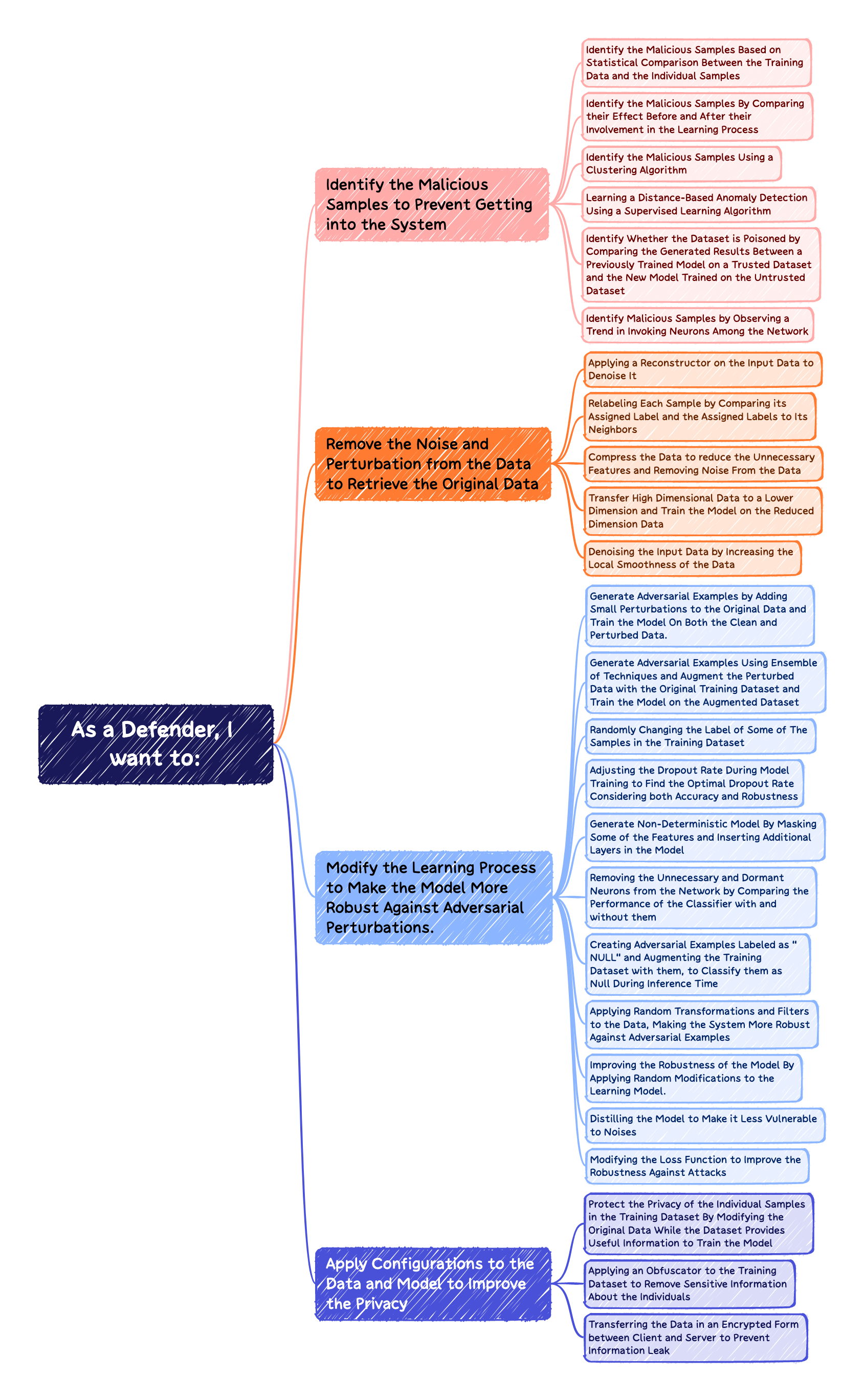}
    \vspace{-30pt}
\vspace*{-139.5ex}  
\begin{center}
   \captionsetup{margin={0\linewidth,180pt}}
    \caption{\textbf{Defensive View of AI/ML ATT\&CK Framework.} Defensive AI/ML Security Workers can explore mitigation scenarios to find the appropriate techniques to practice secure AI/ML design. On the website, the leaf nodes of this view are connected to the concrete mitigation techniques.}
    \label{fig:Defensivescenarios}
\end{center}
\vspace*{138ex}

\end{figure*}

\subsection{How to Use AI/ML ATT\&CK  Framework, Walk-through Examples}
\label{sec:12}

To facilitate using the guidelines of AI/ML ATT\&CK framework, it is delivered as a publicly accessible web solution at \url{http://www.aimlattack.com}.

Using this web solution, users can explore the framework from various perspectives. The following describes how it can be used to support Offensive and Defensive AI/ML Security Workers.

\subsection{Use Case\#1: AI/ML ATT\&CK for Offensive Security Workers} 
An offensive AI/ML Security Worker can use AI/ML ATT\&CK framework to determine the attack surface of an AI-enabled system and deploy offensive techniques to assess the system's robustness. The framework represents \textit{attack scenarios} which have been developed with respect to the \textit{goals of the cyberattack}, \textit{level of access to the target system}, and the \textit{application domain} of the target system. 
By following the provided comprehensive step-by-step decision tree style guidance, as shown in Figure \ref{fig:offensive}, AI/ML Security Workers can identify the related offensive techniques. Furthermore, on the website, by selecting each offensive technique, the user will be directed to a new page that thoroughly explains the technique's details and represents tools capable of implementing the technique to deploy and observe the system's performance under such attack.

\textbf{Example Offensive Scenario:} For a given system that utilizes AI algorithms and has been trained on sensitive data, an offensive AI/ML Security Worker wants to determine to what extent the system is robust against data breaching. Thus, as the first step, by following the offensive view of the framework (Figure \ref{fig:offensive}), the offensive AI/ML Security Worker selects the second box, \textit{``Steal System's Private Information"}. Then, in the next layer, considering the attacker's goal, which is assessing the system's security against data breaching, the AI/ML Security Worker chooses ``Stealing Private Information about the Training Data". Next, with respect to the goal, ``Identify the Participation of a Sample in Training Model" can potentially be chosen. 
Finally, considering the attack's setting, the AI/ML Security Worker can choose any of the nodes of interest (e.g., ``With access to the Model and Training Data"). By selecting the leaf node, the framework will expand all the related concrete techniques and tools.

\subsection{Use Case \#2: AI/ML ATT\&CK for Defensive AI/ML Security Workers} 
The AI/ML ATT\&CK framework empowers defensive AI/ML Security Workers to identify the system's attack surface, reason about the system vulnerabilities, and select the best mitigation techniques to improve the system's robustness. Considering the attacks on the system that AI/ML Security Workers want to mitigate, they can select the best potential mitigation techniques and utilize the provided defensive tools to deploy and improve the system's robustness.

\textbf{Example Defensive Scenario:}
For a given system that utilizes AI algorithms and has been trained on sensitive data, it has been shown that the system is vulnerable to attacks that attempt to identify the participation of individual samples in the training process. Thus, it is required to deploy appropriate mitigation techniques to improve the robustness against such attacks. 
To this end, by following the defensive view represented in the framework, as shown in Figure~\ref{fig:Defensivescenarios}, the user can choose the last box, \textit{``Apply Configurations to the Data and Model to Improve the Privacy"}. In the next layer, the user can select the first box, \textit{``Protect the Privacy of the Individual Samples in the Training Dataset by Modifying the Data while it Provides Useful Information to Train the Model."} Finally, by selecting the leaf node in the website, the framework expands all the appropriate mitigation techniques. Also, by selecting each technique, the user will be directed to a new page that thoroughly explains the details of the mitigation technique, and represents tools capable of implementing the technique. 

\section{User Evaluation}
\label{sec:Evaluation}

A small user study was conducted to show our framework's effectiveness. First, six subject matter experts (SMEs) with prior work experience as AI engineers who were not cybersecurity professionals were recruited. Then, these SMEs were divided into two groups, where the AI experience level of the groups was balanced based on the years of experience.
The groups were provided with the same case study of an Object Detection and Recognition system for an Autonomous Vehicle.
Both groups were asked to identify potential attacks and mitigation techniques across the development cycle of this project. This study had two parts: each user was asked to once play the role of an Offensive AI/ML Security Worker and another time as a Defensive AI/ML Security Worker.
Group \#1 had access to existing taxonomies in the literature (\cite{tabassi2019taxonomy, 10.1145/2046684.2046692, 9099439}) as well as access to the internet, while Group \#2 was requested to use only the AI/ML ATT\&CK framework. For evaluation, we calculated the number of correct answers reported by each user divided by the total number of correct answers (ground-truth).

The results show that Group \#2 identified not only abstract attack classes (types of attacks) but also concrete techniques. The response of Group \#2 was specific, enumerating attack classes, sub-classes, and then listing specific methods from a paper. In contrast, Group \#1 only identified very high-level types of attacks (e.g., poisoning attack) and mitigation techniques and could not be specific about sub-types and concrete methods. In comparison, the responses from Group \#1 lacked information about a concrete method to implement the attack or mitigate it.  
Group \#1 only identified 3\% of concrete attack methods and 5\% of mitigation techniques, while Group \#2 identified 39\% of all correct and concrete attack methods and 50\% of correct mitigation techniques.
Even at the abstract high-level attack classes (types of attacks), Group \#1 identified 75\% of attack types and 69\% of mitigation types correctly, while Group \#2 identified 100\% of both. Overall, Group \#2 outperformed Group \#1 in terms of retrieved correct attacks and mitigation. While the current results are promising, given the sample size, further studies are needed to understand the effectiveness of each part of this framework. 

\section{Limitations and Future Work}
\label{sec:Discussion}
All represented information in this framework has been extracted and characterized through an SLR of the related techniques and tools. However, despite the extensive effort to curate and develop the framework, it is not guaranteed that these techniques and tactics are the only viable ones that can be explored by AI/ML Security Workers, especially with the emergence of new AI/ML cyberattacks.


%
We have only carried out a limited-scope study involving six AI engineers. Further work is required to evaluate the claims related to the framework's effectiveness in supporting AI engineers with limited security experience to improve AI systems' security.

\section{Ethics}
 Offensive cyberwarfare raises serious ethical problems for societies. This topic is not fully addressed by policies and regulations. Our primary intention in formulating the AI/ML ATT\&CK framework and the associated offensive and defensive tools is purely educational and for enhancing the robustness of AI/ML systems. We do not endorse the unethical use of this framework and curated tools/software. Using offensive techniques requires extreme legal and ethical consideration, careful determination of the perpetrators and victims, and reduction of collateral damages.


\bibliographystyle{IEEEtran}      
\bibliography{Main.bib}

\begin{thebibliography}{10}
\providecommand{\url}[1]{#1}
\csname url@samestyle\endcsname
\providecommand{\newblock}{\relax}
\providecommand{\bibinfo}[2]{#2}
\providecommand{\BIBentrySTDinterwordspacing}{\spaceskip=0pt\relax}
\providecommand{\BIBentryALTinterwordstretchfactor}{4}
\providecommand{\BIBentryALTinterwordspacing}{\spaceskip=\fontdimen2\font plus
\BIBentryALTinterwordstretchfactor\fontdimen3\font minus
  \fontdimen4\font\relax}
\providecommand{\BIBforeignlanguage}[2]{{%
\expandafter\ifx\csname l@#1\endcsname\relax
\typeout{** WARNING: IEEEtran.bst: No hyphenation pattern has been}%
\typeout{** loaded for the language `#1'. Using the pattern for}%
\typeout{** the default language instead.}%
\else
\language=\csname l@#1\endcsname
\fi
#2}}
\providecommand{\BIBdecl}{\relax}
\BIBdecl

\bibitem{tabassi2019taxonomy}
E.~Tabassi, K.~J. Burns, M.~Hadjimichael, A.~D. Molina-Markham, and J.~T.
  Sexton, ``A taxonomy and terminology of adversarial machine learning,''
  \url{https://nvlpubs.nist.gov/nistpubs/ir/2019/NIST.IR.8269-draft.pdf}, pp.
  1--29, 2019.

\bibitem{attck}
M.~ATT\&CK, ``{MITRE} {ATT}\&{CK},'' \url{https://attack.mitre.org/}, 2021.

\bibitem{CWE}
MITRE, ``Common weakness enumeration,'' \url{https://cwe.mitre.org/}, 2021.

\bibitem{10.1145/2046684.2046692}
\BIBentryALTinterwordspacing
L.~Huang, A.~D. Joseph, B.~Nelson, B.~I. Rubinstein, and J.~D. Tygar,
  ``Adversarial machine learning,'' in \emph{Proceedings of the 4th ACM
  Workshop on Security and Artificial Intelligence}, ser. AISec '11.\hskip 1em
  plus 0.5em minus 0.4em\relax New York, NY, USA: Association for Computing
  Machinery, 2011, p. 43–58. [Online]. Available:
  \url{https://doi.org/10.1145/2046684.2046692}
\BIBentrySTDinterwordspacing

\bibitem{9099439}
K.~Sadeghi, A.~Banerjee, and S.~K.~S. Gupta, ``A system-driven taxonomy of
  attacks and defenses in adversarial machine learning,'' \emph{IEEE
  Transactions on Emerging Topics in Computational Intelligence}, vol.~4,
  no.~4, pp. 450--467, 2020.

\bibitem{10.1145/3442167.3442177}
\BIBentryALTinterwordspacing
J.~M. Spring, A.~Galyardt, A.~D. Householder, and N.~VanHoudnos, ``On managing
  vulnerabilities in ai/ml systems,'' in \emph{New Security Paradigms Workshop
  2020}, ser. NSPW '20.\hskip 1em plus 0.5em minus 0.4em\relax New York, NY,
  USA: Association for Computing Machinery, 2020, p. 111–126. [Online].
  Available: \url{https://doi.org/10.1145/3442167.3442177}
\BIBentrySTDinterwordspacing

\bibitem{Julie}
J.~Haney and W.~Lutters, ``Cybersecurity advocates: Discovering the
  characteristics and skills of an emergent role,'' \emph{Information \&
  Computer Security}, vol. ahead-of-print, 03 2021.

\bibitem{8805749}
S.~Jia, X.~Liu, P.~Zhao, C.~Liu, L.~Sun, and T.~Peng, ``Representation of
  job-skill in artificial intelligence with knowledge graph analysis,'' in
  \emph{2018 IEEE Symposium on Product Compliance Engineering - Asia
  (ISPCE-CN)}, 2018, pp. 1--6.

\bibitem{8804457}
S.~Amershi, A.~Begel, C.~Bird, R.~DeLine, H.~Gall, E.~Kamar, N.~Nagappan,
  B.~Nushi, and T.~Zimmermann, ``Software engineering for machine learning: A
  case study,'' in \emph{2019 IEEE/ACM 41st International Conference on
  Software Engineering: Software Engineering in Practice (ICSE-SEIP)}, 2019,
  pp. 291--300.

\bibitem{NICE}
R.~Petersen, D.~Santos, M.~Smith, K.~Wetzel, and G.~Witte, ``Nist special
  publication 800-181 revision 1: Workforce framework for cybersecurity (nice
  framework),'' \url{https://nvlpubs.nist.gov/nistpubs/SpecialPublications/
  NIST.SP.800-181r1.pdf}, 2020.

\bibitem{bughin2018skill}
J.~Bughin, E.~Hazan, S.~Lund, P.~Dahlstr{\"o}m, A.~Wiesinger, and
  A.~Subramaniam, ``Skill shift: Automation and the future of the workforce,''
  \emph{McKinsey Global Institute}, vol.~1, pp. 3--84, 2018.

\bibitem{kitchenham2004procedures}
B.~Kitchenham, ``Procedures for performing systematic reviews,'' \emph{Keele,
  UK, Keele University}, vol.~33, no. 2004, pp. 1--26, 2004.

\bibitem{workshop-paper}
\BIBentryALTinterwordspacing
M.~Fazelnia, I.~Khokhlov, and M.~Mirakhorli, ``The limitations of deep learning
  in adversarial settings,'' in \emph{In 9th International Conference on
  Learning Representation, RobustML work- shop, (ICLR), Vienna, Austria
  (Virtual)}, 2021. [Online]. Available: \url{https://arxiv.org/abs/2202.09465}
\BIBentrySTDinterwordspacing

\bibitem{7467366}
N.~Papernot, P.~McDaniel, S.~Jha, M.~Fredrikson, Z.~B. Celik, and A.~Swami,
  ``The limitations of deep learning in adversarial settings,'' in \emph{2016
  IEEE European Symposium on Security and Privacy (EuroS\&P)}, 2016, pp.
  372--387.

\end{thebibliography}

\begin{IEEEbiography}{Mohamad Fazelnia }{\,} is
currently a Ph.D. student at Global Cybersecurity Institute (GCI), Rochester Institute of Technology. His research interest is in the area of AI Security, Human Aspects of Cybersecurity and Software Development. He is a member of the IEEE and the IEEE Computer Society. 

\end{IEEEbiography}

\begin{IEEEbiography}{Ahmet Okutan}{\,} is
a Senior AI Engineer at Global Cybersecurity Institute (GCI), Rochester Institute of Technology. He hold a Ph.D. in Computer Science from Işık Üniversitesi, Turkey. He has over 20 years experience designing, developing and deploying complex software systems. His research and development interest is on the overlap of AI/ML and cybersecurity.   
\end{IEEEbiography}

\begin{IEEEbiography}{Mehdi Mirakhorli,}{\,}(Member, IEEE) is currently an associate professor and Kodak Endowed Chair at Global Cybersecurity Institute (GCI) and Software Engineering Department at Rochester Institute of Technology. He  received his Ph.D. degree in Computer Science from DePaul University, in Chicago. His research interests are in the area of cybersecurity, software assurance and artificial intelligence. He is the Associate Editor for IEEE Transaction on Software Engineering (TSE) and International Journal on Empirical Software Engineering (EMSE).  Contact him at mehdi.mirakhorli@rit.edu.
\end{IEEEbiography}


\end{document}